\documentclass[12pt]{article}
\usepackage{a4}
\usepackage{epsf}

\newcommand{\bea}{\begin{eqnarray}}
\newcommand{\eea}{\end{eqnarray}}

\begin{document}

\begin{center}
\title{To rescue a star}
\author{As. Abada$^a$,  M.B. Gavela$^b$ and
O. P\`ene$^a$} \par
\maketitle
{$^a$ Laboratoire de Physique Th\'eorique et Hautes
Energies\footnote{Laboratoire
associ\'e au
Centre National de la Recherche Scientifique - URA D00063
\\e-mail: abada@qcd.th.u-psud.fr,
gavela@delta.ft.uam.es,
pene@qcd.th.u-psud.fr.}}\\
{Universit\'e de Paris XI, B\^atiment 211, 91405 Orsay Cedex,
France}\\
$^b$Departamento de F\' \i sica Te\'orica, Universidad Aut\'onoma
de Madrid, Canto Blanco, 28049 Madrid.
\end{center}

\begin{abstract}
 Massless neutrinos are exchanged in a neutron star, leading to long range
interactions. Many body forces of this type follow and we resum them. Their
net contribution to the total energy is negligible as compared to the star
mass. The stability of the star is not in danger, contrary to recent
assertions.
\end{abstract}
\begin{flushright} LPTHE Orsay-96/36\\ FTUAM FEV/96/22\\ hep-ph/9605423
\end{flushright}
\newpage

 It has been recently claimed that the multibody exchange of massless
neutrinos renders neutron stars unstable, as the induced self-energy
exceeds the mass of the star \cite{fisc}. Multibody neutron potentials
are generated.  The putative culprits would be the long distance, infrared,
effects associated to neutrino exchange among 4 neutrons or more.
Specifically, the more neutrons
are involved in a given potential, the larger contribution to the  energy
would result. This would be an
exceptional case in physics where the main contribution would not be dominated
by the combinatorial effect of few body potentials.

 Long range forces mean infrared effects. The infrared scale of the problem
is the radius $R$ of the star. Additional parameters of the problem which may
play a significant role are the strength of the weak
 interactions, $G_F$, the neutron density, $n_n$, and a possible neutrino
density, $n_\nu$:

\[ G_F= 1.17\, 10^{-5} \,{\rm GeV}^{-2},\qquad  n_n \sim 0.4\,
{\rm fm}^{-3} \,
\]
 \bea R\sim 10\,{\rm km},\qquad
  n_\nu \sim 4.\, 10^{-23}\, {\rm fm}^{-3} \label{param}  \eea

We first analyze the problem in
the  Hartree-Fock approximation: correlations among neutrons
are neglected. That is, it will be assumed that the neutrons are uniformly
distributed in the star.

\section{Neutrino spectrum in a neutron star}

The properties of neutrinos and antineutrinos that propagate in a medium
differ from those in
the vacuum. In particular, the vacuum energy-momentum relation for massless
fermions, $\omega=|\vec q|$, where $\omega$ is the energy and $|\vec q|$ the
 magnitude of the momentum vector, does not hold in a medium \cite{wolf}.

Assume for the time being that the material of the  neutron star is made
exclusively of neutrons, among which neutrinos are exchanged.
The density-dependent corrections to the neutrino self-energy result, at
leading order, from the evaluation of
$Z^0$-exchange diagrams between the neutrino and the neutrons in the
medium, with the $Z^0$ propagator evaluated at zero momentum. They
can be summarized \cite{nieves} by the following dispersion relation

\bea
\omega=|\vec q|\pm b,
\label{disp}
\eea

where
\bea
b\simeq - \sqrt{2} G_F\, n_n/2 \sim -0.2\, 10^{-7}\,{\rm GeV}\sim -10^{-7}
\,{\rm fm}^{-1}= - 10^{-2}.\,\mbox{\AA}^{-1}
\label{pmb}
\eea

 In eq. (\ref{disp}) the upper(lower) sign refers to neutrinos (antineutrinos).
 $b$ resums the zero-momentum transfer interaction of a massless neutrino
with any number of neutrons present in the media. Sensibly enough, it
depends on the neutron density instead of on the total number of neutrons.

 The dispersion relation shows a displacement of the energy levels for the
different modes, a negative shift for neutrinos, and a positive one for
antineutrinos. The Dirac see level is displaced. Would the neutron star
occupy the whole universe, it
would just mean a change of variables, with no physical consequence.
The finite size of the star changes the picture.
 Notice that $b$ acts as the depth of a potential well \footnote{
When a neutrino-sea is present \cite{loeb},
 the above remarks hold with $b$  given by
$$
  b\simeq \mp \sqrt{2} G_F\, (n_n-n_{\nu})/2
- \frac{8\sqrt{2} G_F \kappa}{3m_Z^2}n_{\nu}<E_\nu>,
\label{b}
$$
where $<E_\nu>$ is the medium average of the neutrino energy, defined in the
rest frame of the medium. These corrections to the fermion propagation would
give higher order effects, and we disregard them in the present paper.
Direct contributions of the neutrino sea to the energy level of the stars
will be considered later, though.}. It is repulsive for antineutrinos
and attractive for neutrinos, which could condense ($b$  plays  also the
role of a chemical potential).

 The effective propagator corresponding to eq.(\ref{disp}) can be written as
\bea
{i\over \widehat q\!\!\! /},
\label{dressed}
\eea

with
\bea
 \widehat q\!\!\! /=(q_0-b)\,\gamma_0\,-\,\vec{q}\,\vec{\gamma},
\label{ay}
\eea
an infrared safe propagator.

 The above results could be obtained as well in an effective lagrangian
approach. Since we are interested in long distance effects, and
correspondingly low momentum exchange, it is a good approximation to use the
following effective lagrangian, as done in ref. \cite{fisc}\footnote{ The
 connection with Fischbach's notation is given by the replacement
 $b=\sqrt 2 G_Fa_nn_n$.},

\bea
{\cal L}_{\mathrm{eff}}(\vec x)=i\,\bar \nu \partial\!\!\! / \,\nu(\vec x) -
 b \bar\nu\gamma_0\nu \ ,
\label{lageff}
\eea
This effective lagrangian is valid inside the star.  The second term summarizes
the
interactions with the neutrons, with the neutron current  reduced in average
to its charge density. For simplicity, we have assumed the   neutron density
to be constant in the star.

 The effective lagrangien is not written in terms of quarks  but in terms of a neutron density 
. Its validity is
thus reduced to a momentum range larger than a typical hadronic size, $\sim 1
\, {\rm fm}$. When treating long distance effects, $\sim 1/R$, this implicit
scale is irrelevant.

\begin{figure}[h]   
\begin{center}
$
\epsfbox{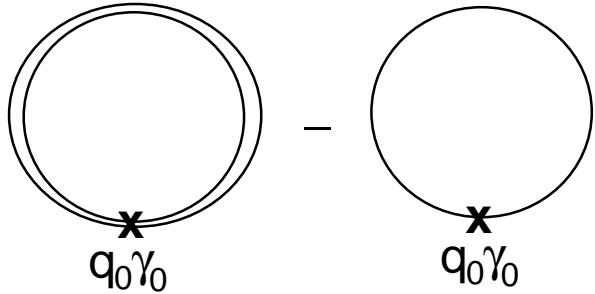}
$
\caption[]{\it{Diagrammatic representation of eq. (8).
The simple line represents the neutrino propagator in real vaccum. The double
line stands for the ``dressed" neutrino propagator, eq (4), which embodies the
interactions with the medium.}} \protect\label{diagram1}
\end{center}
\end{figure}

 Eqs. (\ref{dressed}) and (\ref{ay}) are
the correct solution to the neutrino field equation inside the star which
follows from
eq. (\ref{lageff}).
Since the problematic potential energy under consideration corresponds to
possible long distance divergences,
it is worth to recall a well-known fact: eqs. (\ref{dressed}) and (\ref{ay})
can be derived as well in perturbation by summing $1,2.......\infty$
interactions of an ``undressed" neutrino with the neutrons. In the infrared,
for
$|\vec q|< b$, this perturbative method fails, though, because
the infinite sum does not converge, it is infrared divergent. The derivation
of  eqs. (\ref{dressed}) and (\ref{ay}) from eq. (\ref{lageff}) has no such
limitation. It has incorporated the interactions with the media, including the
long distance ones. This suggests that no long distance divergence is to be
expected from physical effects stemming from the interior of the star. We
could stop the argument here. For the sake of comparison with
previous literature, we explicit the computation below.

\section{ Energy density from neutrino exchange}

A traditional way to estimate the energy induced by neutrino exchange is to
compute first the possible $2,3,4, ...$ exchange potentials and then  add their
contributions integrated over the neutron positions in the star. We  use
 instead  a simpler and more direct method, formally equivalent to the latter,
 as proven in section 4.

 As remarked by Fischbach, eq. (B2) in \cite{fisc}, Schwinger has
 provided the tools to
 compute the density of weak interaction energy, $w$, due to neutrino
 exchanges. It is
 given by  the  difference between the energy density for a neutrino
 propagating in the ``vacuum'' defined by the neutron star,
 $|{\hat{0}}>$, and the corresponding one for the real, matter-free, vacuum,
 $|0>$,

\bea
 w=<\hat{0}|{\cal H}(0)|{\hat{0}}> - <{0}|{\cal H}_0(0)|{0}>,
\label{energy}
\eea
where ${\cal H}_0(0)$
is the free hamiltonian density for the propagating neutrino, and
 ${\cal H}(0)$
includes the interaction with the neutrons, as given by eq. (\ref{lageff}).

 In diagramatic form, depicting the ``dressed'' massless fermion propagator,
eq. (\ref{dressed}), by a double line,  the formal eq. (\ref{energy})
 corresponds to the computation of the diagrams in fig. 1,

\bea
\int \frac{d^4q}{(2\pi)^4}(-i) {\rm Tr}\left[q_0\gamma_0\left(
{1\over \widehat q\!\!\! /}-{1\over q\!\!\! /}\right )L
 \right] ,\qquad L= \frac {1-\gamma_5}2.
\label{design}
\eea

 This is a formal expression. A regularization procedure must be chosen. Let
us consider Pauli-Villars. The problem has no
infrared divergences, and it will control ultraviolet ones.
Upon Pauli-Villars regularization, we can make a shift of variables, and the
total result is zero, using the symmetry $q_\mu \to - q_\mu$.

\begin{figure}[h]   
\begin{center}
$
\epsfbox{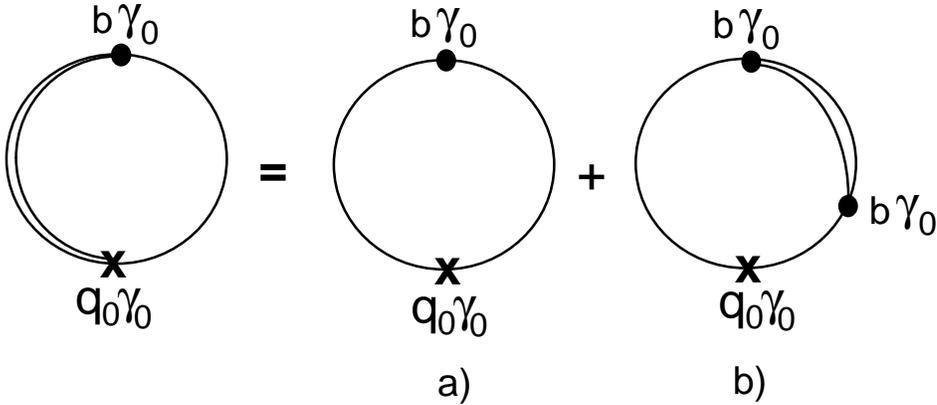}
$
\caption[]{\it{The diagrams in fig. (\ref{diagram1}), developped  according to
Schwinger-Dyson. See eq. (10).}} \protect\label{diagram2}
\end{center}
\end{figure}

We can gain further insight by using the Schwinger-Dyson expansion,

\bea
{1\over \widehat q\!\!\! /}={1\over  q\!\!\! /} +{1\over  q\!\!\! /}\ b
\gamma_0\
{1\over \widehat q\!\!\! /},\label{SD}
\eea

and eq.(\ref{design}) is equivalent to the computation of the diagrams in
fig. 2,

\[ w=
\int \frac{d^4q}{(2\pi)^4}(-i) {\rm Tr}\left[q_0\gamma_0\,
{1\over  q\!\!\! /}b \gamma_0\,{1\over \widehat q\!\!\! /}\,L
 \right]=\int \frac{d^4q}{(2\pi)^4}(-i) {\rm Tr}\left[q_0\gamma_0\,
{1\over  q\!\!\! /}b \gamma_0\,{1\over  q\!\!\! /}\,L
 \right]\,+\]
\bea
\int \frac{d^4q}{(2\pi)^4}(-i) {\rm Tr}\left[q_0\gamma_0\,
{1\over  q\!\!\! /}b \gamma_0\,{1\over \widehat q\!\!\! /}
  b \gamma_0\,{1\over  q\!\!\! /}º,L \right]\,.
\label{design2}
\eea

The first term on the right hand side of this equation corresponds in fact to
 the self-energy of one neutron, and strictly speaking it should be
substracted (upon Pauli-Villars regularization, it gives a null contribution
 by itself, though). We see then that the neutrino contribution to the
energy density
stems formally from the computation of the diagram b) in fig.(2), that is,

\bea  w=
\int_{PV} \frac{d^4q}{(2\pi)^4}(-i) {\rm Tr}\left[q_0\gamma_0\,
{1\over  q\!\!\! /}b \gamma_0\,{1\over \widehat q\!\!\! /}
  b \gamma_0\,{1\over  q\!\!\! /}\,L \right]\,=\, 0,
\label{bondia}
\eea
where $PV$ stands for Pauli-Villars.
The result holds in any regularization
  scheme invariant for shifts of the energy variable.

\section{Energy density from neutrino condensate}

 Before turning to the connection with Fischbach's work and the discussion
of the regularization procedure, we should consider the presence of a neutrino
sea inside the neutron star. This will be related to finite size effects.
 Indeed, Smirnov and Vissani have recently recalled that such a sea is
present \cite{loeb}, and they argued that the two-body
potential in the star is modified \cite{smirnov}
 due to the blocking effects of the sea,
damping the long range forces.

 We are interested in resumming the contribution to the energy
from multibody exchange, to all orders. In our formalism this is easily done.
It suffices to replace in eqs. (\ref{design})-(\ref{bondia}) the dressed
 propagator, eq.(\ref{dressed}), by

\bea
\widehat S_{\mathrm F} = i\left ( {1\over \widehat q\!\!\! /} + 2 \pi i\
\widehat q\!\!\! / \ \delta(\widehat q^2) \theta(-q_0)\theta(q_0-b)\right )\,,
\label{propagateurH}
\eea
where the Fermi sea contains the neutrino states that have a negative
energy in the star, i.e., according to eq. (\ref{disp}), $|\vec q|<|b|$ or
 $b<q_0<0$. As stated in \cite{loeb} these neutrinos are trapped inside
the star by the attractive potential while
the antineutrinos are repelled away from the star.

Denoting by $w_s$ this new contribution to the energy density, it follows that
\[ w_s=
\int \frac{d^4q}{(2\pi)^4}(-i) {\rm Tr}\left[q_0\gamma_0\,
{1\over  q\!\!\! /}b \gamma_0\,2\pi i {\widehat q\!\!\! /}
\delta(\widehat q\,^2)\theta(-q_0)\theta(q_0-b)\,
  b \gamma_0\,{1\over  q\!\!\! /}\,L \right] =\]
\bea
 +\int \frac {d^3q}{(2\pi)^3}
(b+|\vec q|)\theta(-b-|\vec q|)=- \frac{b^4}{24\pi^2}\sim -   10^{-31}
\,{\rm GeV\, fm}^{-3}.\label{mer}
\eea

 This equation has an inmediate transparent interpretation: it is the
contribution to the total energy density due to the neutrino Fermi-sea, which
 is supposed to be filled up in a neutron star. Quantitatively, a tiny one.
Its physical meaning stems
from the finite size of the star, that is, from the difference of the energy
levels between the dense medium and the free vacuum. In other words, would the
neutron star occupy the whole universe, antineutrinos could not have been
expelled out through its surface. Indeed, it is easy to check that $w_s$ would vanish if a
term representing an antineutrino sea would be added to
 the propagator, eq. (\ref{propagateurH}).

 Given the result in the previous section, the total energy density, eq.
(\ref{energy}), is then given by

\bea
w=w_s\sim -10^{-31}\,{\rm GeV\, fm^{-3}} .
\eea

\section{Comparison with Fischbach's analysis}

 In order to compare with Fischbach analysis \cite{fisc}, eq.(\ref{bondia})
can be developed using the Schwinger-Dyson equation (\ref{SD}).
 This leads to

\bea w=
\sum_{k=2}^\infty
\int \frac{d^4q}{(2\pi)^4}(-i) {\rm Tr}\left[q_0\gamma_0\,
{1\over  q\!\!\! /}\left(b \gamma_0\,{1\over  q\!\!\! /}\right)^k\,L\right].
\label{expansion}
\eea

Each  term in the series of eq. (\ref{expansion}) correspond exactly to
those considered by
Fischbach, with the difference that the sum does not stop at the total number
of neutrons in the star, but runs up to $\infty$. It should be so: a
 given neutrino can interact several times with the same neutron, a fact
not included in Fischbach treatment.

He has estimated the $k-$body potential
for non-zero momentum transfer, and then integrated over the relative
positions of the neutrons. This is equivalent
to computing the loop in eq. (\ref{expansion}) in which all of the
 $k$ neutron current insertions
 have vanishing momentum transfer. With the method used in this paper,
 an insertion
of the neutrino energy operator, $q_0 \gamma_0$, is present as well. Upon
 integration by parts\footnote{This integration by parts corresponds to
 the steps described in eqs. (B27)-(B29) in \cite{fisc}. }
with a proper regularization,
 it is straightforward to rewrite it as

\bea
w=\sum_{k=2}^\infty
\frac 1 k \int \frac{d^4q}{(2\pi)^4}(-i) {\rm Tr}\left[
\left(b \gamma_0\,{1\over  q\!\!\! /}\right)^k\,L\right].
\eea
The factor $1/k$ appearing in this formula is the one required to obtain
 exactly the expansion of the logarithm, eq. (B34), in \cite{fisc}.

 Taken separately, each term in the  expansion (\ref{expansion})
 with 4 or
 more neutron insertions,
$k\ge 4$, is infrared divergent, the degree of divergence rising with the
number of neutron insertions. They would lead to a contribution
to the energy density
of the form $b^4 (b\,R)^{k-4}$, where $b\,R\sim 1. 10^{12}\gg 1$. This
reasoning lead Fiscbach to conclude that neutron stars are unstable if the
neutrinos are massless\footnote{Smirnov and Vissani, \cite{smirnov} revised
 version may 24th, argue that due to Pauli blocking, the infrared dangerous
 parameter $O(bR)$ in Fischbach's expansion is replaced by a parameter $O(1)$
 and the series might be resummed. We have performed the sum with (section 3),
or without (section 2) Pauli blocking and find no infrared divergence.
}. Nevertheless, as remarked above, the correct behaviour
which follows from the field equations for the lagrangian, eq.(\ref{lageff}),
proves that the sum of the series is infrared convergent, as evident from
eq. (\ref{bondia}). We are facing a situation where the behaviour of
 the different
 terms in a perturbative expansion differs essentially from the resummed,
non-perturbative, result\footnote
{From an algebraic point of view
the behavior  is similar to that of
the function $1/(1-x)$, for $x >1$, which is wellfined even if its
 expansion in $x$ diverges.}. { \it Multibody massless neutrino
exchange do not destabilize the star}.

 For the sake of comparison with Fischbach's paper it is worth to comment
 upon the treatment of ultraviolet divergences, even if the latter are
irrelevant to
 the long distance problem under discussion, as already remarked in
\cite{fisc}. As finally such a problem does
not exist, they could be the leading contribution to the star
 self-energy, although a harmless  one.

  Fischbach has correctly remarked
that only terms with 4 or less neutron insertions, $k \le 4$, are ultraviolet
 divergent.
He explicitly computed their finite contribution to the star self-energy.
 For $k=2$ he obtained a non-zero result, contrary to our results
 above. The issue is not the form of the 2-body exchange potential,
 but the energy density derived from it upon spatial integration,
where the ultraviolet divergence appears. The difference
stems from the regulator choice. For simplicity, we have
chosen above to work with Pauli-Villars, a perfectly consistent one.
Fischbach implemented a
 cutoff non invariant for the shift of energy variables,
 a harder one, leading to a finite result. He points out that there is a
natural ultraviolet cutoff in the problem - the ``hard core'' $r_c$,

\bea
r_c \simeq 0.5\, \, { \rm fm} ,
\eea
which prevents neutrons from ``piling up'' in space. It can be interpreted
as the natural hadronic cutoff for the effective lagrangien,
eq. (\ref{lageff}), discussed above. To consider the effects of such a
cutoff is tantamount to go beyond the Hartree Fock aproximation, where
correlations among neutrons are neglected. $r_c$ is then to be added to
our list of physically relevant parameters in eq. (\ref{param}).
 Its exact  shape is not known from first principles. We do not wish to discuss
 here Fischbach's choice, as it only affects the multibody potentials
involving  4 neutrons or less, irrelevant for the infrared problem under
discussion. The physical consequences, if present, should be derivable from
the underlying
 theory, the standard model. We do not follow this line of research here, as
the
main focus of this paper are long distance effects, for which $r_c$ is
 irrelevant.
 His result, even if larger
than the  contribution of the Fermi sea derived here, does
not threaten the stability of the star. Would we use Fischbach's cutoff,
our dominant contribution would be provided by the term with the strongest
 ultraviolet divergence, $k=2$: from two body
 neutrino exchange. The result is, eq. (4.2) in\footnote{We neglect
 corrections of order $1/(R^3\,n_n)\sim 10^{-57}$.}
 \cite{fisc},
\bea
 {w^{(2)}}= \frac {b^2}{8\pi^2 r_c^2}\sim 10^{-16}\, {\rm GeV\, fm}^{-3},
\label{w22}
\eea
to be compared to the neutron mass density $\sim O(1$ GeV fm$^{-3})$. Eq.
(\ref{w22}) does not take into account the presence of the neutrino sea. As
 the latter only corrects the long distance behaviour, the
minor changes resulting from its inclusion do not change the physical result
 for the star energy density.

\section{Conclusion}

 We have shown that the resummation of multibody neutrino exchange in a
neutron star results in an infrared well-behaved contribution to the star
energy density. This holds whether  the presence of a neutrino
sea is taken into account or not.  The resulting energy density is negligible
as compared to the star mass density.
 The star remains stable, even if neutrinos are massless.

\section*{Acknowledgements.}

This work was supported in part by the Human Capital
and Mobility Programme, contract CHRX-CT93-0132 and through funds
 from CICYT, project AEN93-0673. O. P\`ene thanks the BBV fundation
for support. We acknowledge Jean-Pierre Leroy
for several inspiring discussions. We are specially indebted to Alvaro De
R\'ujula for important discussions and comments.

\end{document}